\documentstyle[twocolumn,aps]{revtex}
\newcommand{\be}{\begin{equation}}
\newcommand{\ee}{\end{equation}}
\newcommand{\bea}{\begin{eqnarray}}
\newcommand{\eea}{\end{eqnarray}}

\begin{document}
\bibliographystyle{prsty}
%LOCALIZED SPATIO-TEMPORAL CHAOS IN PARAMETRICALLY EXCITED WAVES\\
\title{Phase Diffusion in Localized Spatio-Temporal Amplitude Chaos}
\author{Glen D. Granzow and Hermann Riecke}
\address{Department of Engineering Sciences and Applied Mathematics,
Northwestern University, Evanston, IL 60208, USA}
\maketitle

\begin{abstract}
We present numerical simulations of coupled Ginzburg-Landau equations
describing parametrically excited waves which reveal persistent dynamics due to
the occurrence of phase slips in sequential pairs, with the second phase slip 
quickly following and negating the first. Of particular interest are 
solutions where these double phase slips occur 
irregularly in space and time within a spatially localized region.
An effective phase diffusion equation utilizing the long term phase 
conservation of the solution explains the localization of this 
new form of amplitude chaos.
\end{abstract}

\centerline{\today}

Spatio-temporal chaos poses one of the important challenges in the investigation of
spatially extended dynamical systems. 
A central question regarding these systems concerns the characterization 
and classification of the chaotic states.
This has been studied most extensively for chaotic traveling
waves in one dimension within the framework of a complex Ginzburg Landau
equation.
There, two different dynamic regimes have been identified, a phase-chaotic
regime and an amplitude-chaotic regime \cite{ShPu92,EgGr95,ChMa96,MaCh96}.
The latter is characterized by the occurrence of phase slips during which the
amplitude of the wave goes to zero and the total phase of the system changes 
by $2 \pi$, i.e. a wavelength is inserted or eliminated.
In the phase-chaotic regime essentially no phase slips occur \cite{EgGr95,MaCh96}. 
%and the amplitude never goes to zero.
This chaotic regime can be described by an equation for the phase
of the wave alone \cite{Ku78}.
In the amplitude-chaotic regime this is not possible since the phase equation
breaks down during phase slips.

In this letter we present a different class of spatio-temporal chaos, which we
have identified in one-dimensional simulations of parametrically excited waves. 
The observed chaos is characterized by the occurrence of {\it double 
phase slips} which effectively preserve the total phase.
Thus, although the phase description breaks down during each of the phase slips,
the long-time dynamics can still be described by an {\it effective} phase 
equation.
The second striking result of our simulations is the localization of 
spatio-temporal chaos in a part of the homogeneous system.
That is, there are solutions that include a spatial region, 
bounded by quiescent regions on each side, in which the 
dynamics are irregular in space and time.
Experimentally, similar phenomena have been
observed in Taylor vortex flow \cite{BaAn86},
Rayleigh-Benard convection \cite{CiRu87}, and parametrically excited surface
waves \cite{KuGo96}.
So far, the localization mechanism in these systems is only poorly understood.

For the system of equations discussed in this letter
(which model parametrically excited waves)
the localization can be understood using the effective phase equation.
We confirm the validity of this approach 
by explicitly determining the effective phase 
diffusion coefficient.
The short-scale chaotic dynamics amount to fluctuations on the large scales captured by
the phase decription.  They may therefore introduce
a noise term in the effective phase diffusion equation similar to that
arising in the
Kardar-Parisi-Zhang equation \cite{KaPa86}, which describes the large-scale behavior of
spatio-temporal chaos in the  
Kuramoto-Sivashinsky equation \cite{Ya81,Za89,JaHa93}, 
the leading-order phase equation for the complex Ginzburg Landau equation \cite{Ku78}.

Parametrically excited waves arise quite generally in systems exhibiting 
weakly damped modes that are oscillatory in both space and time, when these 
modes are forced at twice their natural frequency \cite{RiCr88,Wa88}.
Faraday's experiment in which surface waves are formed on a liquid in a 
vertically oscillated container is perhaps the best known example 
(e.g. \cite{CrHo93,TuRa89,KuGo96}).
%\cite{Fa31}.
Other examples are parametrically forced 
spin waves \cite{Su57,El93}, and traveling waves in electroconvection 
\cite{ReRa88} and in Taylor Dean flow \cite{TeAn96}.
%Surprisingly, they have even been found in Faraday experiments with granular
%media instead of fluids \cite{MeUm95}.
For small forcing amplitude, these systems can be modeled by a pair of coupled 
Ginzburg-Landau equations \cite{RiCr88,Wa88},
\begin{eqnarray}
\partial_TA+s\partial_XA &=& d\partial_X^2A+aA+bB \nonumber \\
& & +cA(|A|^2+|B|^2)+gA|B|^2\\
\partial_TB-s\partial_XB &=& d^*\partial_X^2B+a^*B+bA \nonumber \\
& & +c^*B(|A|^2+|B|^2)+g^*B|A|^2.
\end{eqnarray}
The dependent variables $A$ and $B$ are complex, and represent 
amplitudes of left and 
right traveling waves which are summed together to yield the solution of the 
underlying system:
\begin{eqnarray}
u(x,t) &=& \epsilon A (X,T)e^{i(q_cx-\frac{\omega_e}{2}t)}
          +\epsilon B (X,T)e^{i(q_cx+\frac{\omega_e}{2}t)} \nonumber \\
       & & +c.c.+O(\epsilon^2).
\end{eqnarray}
They vary on slow time and space scales, $T=\epsilon^2t$ and $X=\epsilon x$ 
respectively.
%The real part of the coefficient of the linear term $a$ gives the linear 
%damping of the (unforced) traveling waves, while the imaginary part of $a$ 
%gives the 
%difference between their natural frequency and half the forcing frequency
%$\omega_e$.
The coefficient $b$ of the linear coupling term gives the amplitude of the 
periodic forcing.

Stationary solutions to equations (1,2) of the form 
\( A = A_0 e^{iqX} \), \( B =B_0 e^{iqX} \) 
(where $A_0$ and $B_0$ are complex constants) 
include those where the amplitudes of $A$ and $B$ 
are equal, corresponding to standing waves in the underlying system
which are phase-locked to the external forcing
%, and solutions where their amplitudes are not equal, 
%corresponding to traveling waves in the underlying system 
\cite{RiCr88,Wa88}.
In this letter we discuss results of numerical simulations of the ensuing
dynamics when the standing waves are perturbed.
These simulations were performed using a Fourier spectral method
in space (periodic boundary conditions) and a fourth order 
Runge-Kutta/integrating-factor scheme in time.

%Close to onset (that is, for forcing amplitudes very close to the minimum value 
%at which the standing waves are excited) equations (1,2) can be 
%reduced to a single Ginzburg-Landau equation with real coefficients 
%\cite{Ri90a}. 
%This implies that the waves are stable near onset in a band of wavenumbers
%which is limited by the Eckhaus instability.
%Near the minimum (which represents onset) the neutral and Eckhaus curves are 
%parabolic.
%For larger forcing amplitudes $b$, the 
%Eckhaus curve need no longer be parabolic and can assume a variety of shapes.
%For the parameters chosen for this study, the Eckhaus curve closes on itself 
%as shown in figure 1 \cite{Ri90a}.
%Thus for $b>0.85$ all standing waves become unstable, somewhat reminiscent of
%the Benjamin-Feir instability of traveling waves \cite{BeFe67,Ne74}.
 
The linear stability diagram for the standing-wave solution of equations (1,2)
with the parameters chosen for this study is shown in figure 1 \cite{Ri90a}.
For small values of $b$ ($b=0.1$, say) solutions to equations (1,2) behave like those of
a single 
real Ginzburg-Landau equation; given an initial condition slightly to the right 
of the Eckhaus stable region, the solution will undergo a phase slip which reduces the 
wave number and moves the solution into the stable region 
(e.g. \cite{KrSc88}).
This single-phase-slip behavior is shown in figure 2a.
For larger values of $b$ however ($b=0.6$, say), the behavior of the solution is 
markedly different;
given an initial condition slightly to the right of the Eckhaus stable region, the 
solution undergoes a phase slip which reduces the average wave number but a 
short time later undergoes a second phase slip at essentially the same location 
which restores the wavenumber to its original value.
This double phase slip, shown in figure 2b, causes the solution to remain 
in the Eckhaus unstable region, thus allowing persistent dynamics.

\begin{figure}[htb]
\begin{picture}(420,180)(0,0)
\put(-50,-65) {\includegraphics{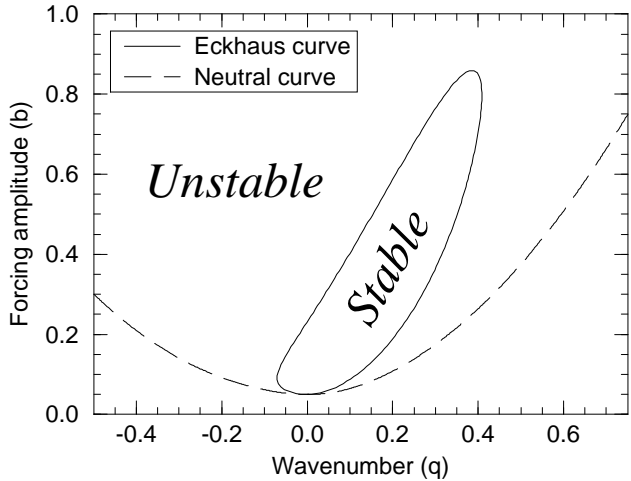}}
\end{picture}
\caption{Stability diagram for parametrically driven standing waves with 
$a=-0.05$, $c=-1+4i$, $d=1+0.5i$, $s=0.2$, $g=-1-12i$.
\protect{\label{f:fig1}}
}
\end{figure}

Simulations of a small system (5 wavelengths long) show that the complexity of 
the ensuing dynamics is dependent on the distance of the initial condition from 
the Eckhaus stable region.
Very close to the Eckhaus curve simple periodic solutions occur.
In these solutions double phase slips occur at only one 
location in space, periodically in time.
%The location depends on the initial conditions, which were chosen 
%close to a stationary solution
%but perturbed such that the local wavenumber contains a maximum which will 
%trigger the first phase slip.
%The period of these solutions diverges as the initial condition is 
%moved into the Eckhaus stable region.
%Just inside the Eckhaus curve there is a small range of wavenumbers wherein 
%both the stationary solution and the periodic solution are stable.
%Further inside the Eckhaus stable region the system behaves like an excitable 
%medium; 
%a sufficiently large perturbation causes a single double-phase-slip
%before the solution relaxes to its final stationary state, with the size of 
%the perturbation required increasing as the initial state is moved further into 
%the stable region.
Further to the right, away from the Eckhaus stable regime
the simple periodic solutions lose stability to more complicated 
periodic solutions and these more complicated periodic solutions in turn lose 
stability to solutions where the double phase slips occur irregularly in 
time and space over the entire domain.
This irregular occurrence of double phase slips is a form of 
spatio-temporal amplitude chaos.
%While we did find a single period doubling (where the location of the double 
%phase slips alternated between two locations in space) when the simple periodic 
%solution lost stability, we did not observe a period doubling cascade 
%[REFERENCE] as the route to chaos.
%as the route to chaos.
Starting with an initial condition far to the right of the Eckhaus curve, 
single phase slips occur which reduce the wavenumber and thus move 
the system to the left, into a region where a stationary, periodic, or chaotic 
solution is stable.

\begin{figure}[htb]
\begin{picture}(420,195)(0,0)
\put(-50,-65) {\includegraphics{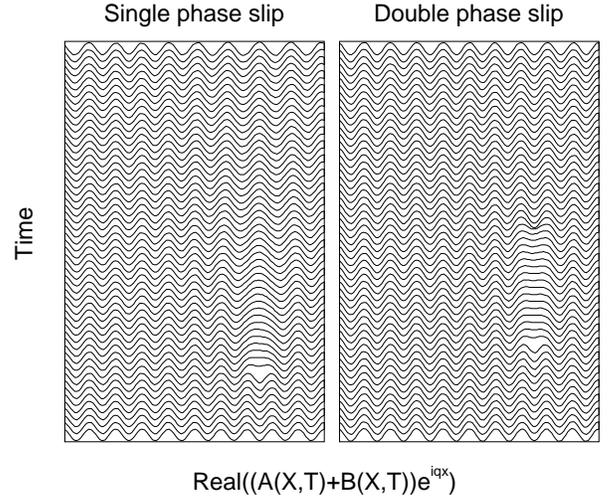}}
\end{picture}
\caption{Space-time diagrams illustrating a) single phase slip observed
for small values of the forcing amplitude (e.g. $b=0.1$), and b) double
phase slip observed for larger values of the forcing amplitude (e.g. $b=0.6$).
Other parameters as in figure 1.
\protect{\label{f:fig2}}
}
\end{figure}

In larger systems (here 47 wavelengths long) and at intermediate distances from
the Eckhaus stable regime we find that the chaotic activity need not spread 
through the whole system.
Instead, double phase slips occur only in a spatially confined region.
This is illustrated in figure 3 which
shows the location in space and time of double phase slips.
Here an initial maximum in the local wavenumber triggers a double phase slip 
which subsequently
causes more double phase slips in the neighborhood of the first.
These double phase slips occur irregularly in space and time within a region 
whose size initially grows but remains bounded in space.
Outside of this region the solution remains stationary (i.e. in physical
space there are regular, periodic standing waves, cf.(3)).

\begin{figure}[htb]
\begin{picture}(420,180)(0,0)
\put(-40,-65) {\includegraphics{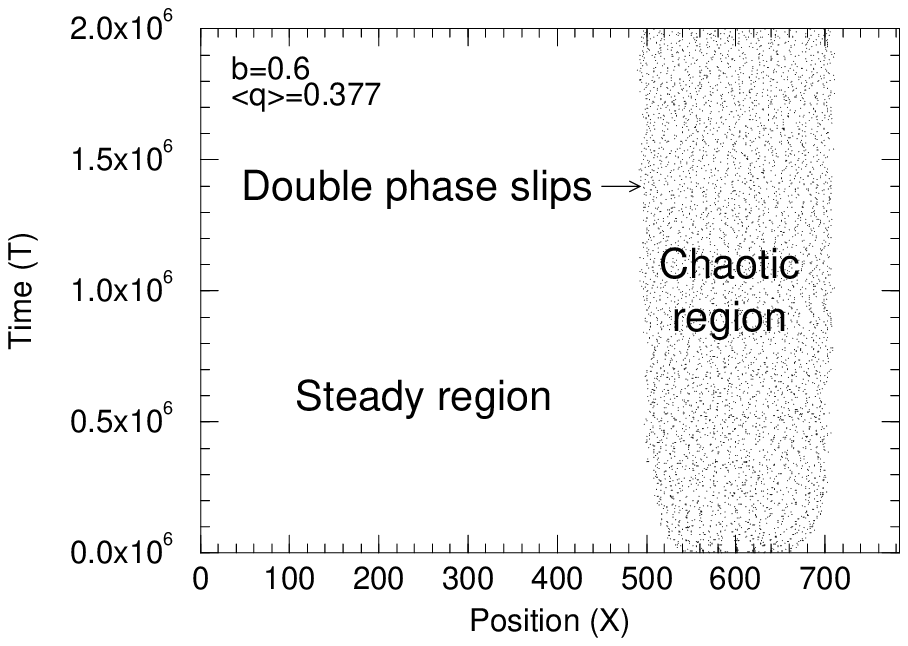}}
\end{picture}
\caption{Space-time diagram showing each double phase slip as a single dot. 
The spatio-temporal chaos is confined to a small part of the system
($b=0.6$, other parameters as in figure 1;  
the averaged effective wavenumber is $<q>=0.377$).
\protect{\label{f:fig3}}
}
\end{figure}

To understand the localization mechanism we turn to the phase diffusion equation
\be
\partial_T \phi (X,T) = D(q) \partial_{X}^{2} \phi (X,T), \label{e:phde}
\ee
which describes the slow evolution of the phase $\phi$ of a steady pattern. 
%The local wave number $q$ is related to the phase $via$ $q=\partial_X \phi$.
Equation (4) is nonlinear through the dependence of the diffusion coefficient
on the wave number $q \equiv \partial_X \phi$.
The use of the phase equation is motivated by the observation that the 
temporally averaged local 
wavenumber is larger in the chaotic region than in the steady region.
The state shown in figure 3 can therefore be considered to consist of two 
domains, one with large and one with small wavenumber.
It has been shown previously \cite{BrDe89,Ri90a,RaRi95a} that such domain structures can 
be stable if the diffusion coefficient $D(q)$ is negative (indicating an unstable
solution) only over a small range of wavenumbers $q$.
Initial conditions with a uniform wavenumber in this range will then evolve to a 
structure consisting of domains in which the local wavenumber lies in either of
the two adjacent ranges of stable wavenumbers.
The stability of the domain structure is due to the 
conservation of the total phase and the instability of the state with
the corresponding uniform wavenumber.

As a basic condition for this mechanism to explain the localization shown in 
figure 3 the phase
description has to be valid, requiring in particular that the phase is conserved.
Across individual phase slips the phase is not conserved
and the phase equation (4) breaks down.
However, across a {\it double} phase slip the phase {\it is} conserved.
On sufficiently long time scales we therefore expect an {\it effective} phase 
equation to be appropriate. Reflexion symmetry in space but not in time
suggests an equation of the form (\ref{e:phde}) with the phase and the diffusion 
coefficient replaced by an effective phase $\hat{\phi}$ and an
effective diffusion coefficient $\hat{D}(\hat{q})$.
To substantiate this claim we determine $\hat{D}(\hat{q})$ by measuring the 
(diffusive) response of the extended chaotic state to a localized time-periodic forcing.
Specifically, we include a spatially local, time-periodic advection 
term in equations (1,2)
%by replacing $s$ in (1) and (2) by $s+v$ and $s-v$ respectively, where
%\begin{equation}
%v = v_{0}\sin(\omega_0 T) \mbox{\hspace{1ex} for $X_1 \leq X \leq X_2$}
%\end{equation}
giving:
\begin{eqnarray}
\partial_TA+(v+s)\partial_XA &=& d\partial_X^2A+aA+bB \nonumber \\
& & +cA(|A|^2+|B|^2)+gA|B|^2\\
\partial_TB+(v-s)\partial_XB &=& d^*\partial_X^2B+a^*B+bA \nonumber \\
& & +c^*B(|A|^2+|B|^2)+g^*B|A|^2\\
%v &=& v_{0}\sin(\omega T), X_{1} \leq X \leq X_{2} \\
%v &=& 0, X < X_{1}, X > X_{2}
\mbox{with $v$} &=& \left\{ 
\begin{array}{ll} 
v_{0}\,\sin(\omega T) & \mbox{for $X_1 \leq X \leq X_2$} \nonumber \\
   0                  & \mbox{otherwise.} \nonumber 
\end{array}
\right.
\end{eqnarray}
%and $v=0$ otherwise.
For relatively small values of $\omega_0$ formerly stationary solutions start to drift 
within the region $X_{1} \leq X \leq X_{2}$, periodically reversing their direction
according to the sign of $v$.
For the regions $X < X_{1}$, and $X > X_{2}$ the situation resembles an 
imposed boundary
condition at $X = X_{1}$, and $X = X_{2}$ at which the phase of the solution
varies sinusoidally.
This is similar to the approach used in experiments
on turbulent Taylor vortex
flow wherein an endcap was moved sinusoidally \cite{WuAn92}.

\begin{figure}[h]
\begin{picture}(420,180)(0,0)
\put(-50,-65) {\includegraphics{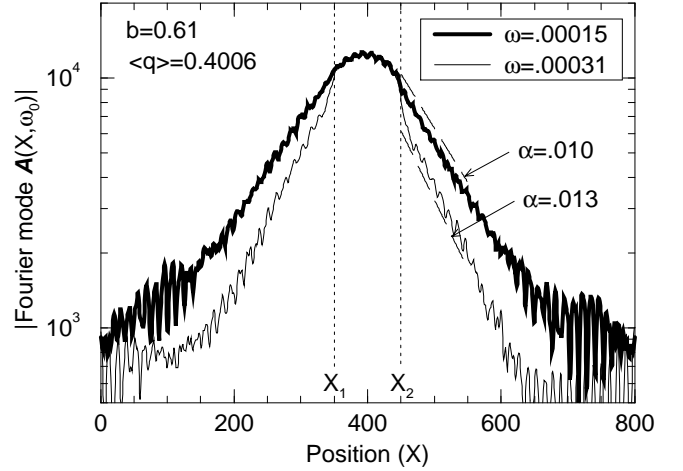}}
\end{picture}
\caption{Decay of the magnitude of the coefficient of the time Fourier transform
of A corresponding to the frequency $\omega_0$ in the localized advection term $v$ 
%given by equation (5),
in equations (5,6),
for two different values of $\omega_0$ ($b=0.61$, other parameters as in figure 1;
the averaged effective wavenumber is $<q>=0.4006$).
\protect{\label{f:fig4}}
}
\end{figure}

The solution to the diffusion equation 
$\partial_t \hat{\phi} (X,t) = \hat{D} \partial_{X}^{2} \hat{\phi} (X,t)$ for $X>X_2$, with 
$\hat{\phi} (X_2,t) = \hat{\phi}_0\, \sin(\omega_0 t)$, is 
\begin{eqnarray}
\hat{\phi} (X,t) &=& \hat{\phi}_0 e^{- \alpha (X-X_2) } \sin(\omega_0 t - \beta (X-X_2)) \\
\mbox{where $\alpha$} &=& \beta = \sqrt{\frac{\omega_0}{2\hat{D}}}.
\end{eqnarray}
If the phase of the solution $A$ in the region $X > X_{2}$ obeys this 
relationship, then the magnitude of the
Fourier mode ${\cal A}(X,\omega_0)$ 
corresponding to the frequency $\omega_0$ 
%in the advection term
will decay exponentially in space as X increases from $X_2$.
Figure 4 shows that this exponential behavior is realized in our simulations.
Note that the data for figure 4 is from a simulation that produced double phase
slips irregularly in space and time over the entire domain (not 
localized chaos).
The phase slips introduce noise which can be seen in figure 4. 
In order to reduce the effect of this noise and extract a reliable decay rate 
$\alpha$, the Fourier integral has been extended 
over many periods (88) of the function $v(X,T)$ in the advection term 
in equations (5,6).
%(5).
Equation (8) shows that for diffusive behavior the decay rate $\alpha$ is 
proportional to $\omega_0^{1/2}$.
The ratio of the slopes of the curves in figure 4 are consistent with this diffusive scaling.
%Note that for larger frequencies and therefore larger gradients of $\hat{\phi}$
%higher order terms  
%(like $-\hat{G} \partial_X^4 {\hat \phi}$, etc.) 
%have to be kept in the diffusion equation which affect the dependence on 
%$\omega$.
%For $\hat{G}>0$, $\alpha$ is therefore expected to go to zero slower 
%than $\omega^{1/2}$.
%\be
%\frac{\alpha(\omega_1)}{\alpha(\omega_2)} = 
%\left(\frac{\omega_1}{\omega_2}\right)^{1/2}[1+\frac{\hat{G}}{2\hat{D}^2}
%(\omega_2-\omega_1)+ O(\omega_1\omega_2)].
%\ee
%We found analogous behavior in the non-chaotic regime where our simulations
%were able to reproduce the analytical value of the diffusion coefficient.
We conclude that the dynamics are indeed diffusive and show in 
figure 5 the diffusion coefficient as a function of the wavenumber.
The solid curve shows the analytical phase diffusion coefficient for the 
stationary
solution \cite{Ri90a} while the triangles show the effective phase
diffusion coefficient for the chaotic solution as measured using the 
decay rate $\alpha$ of the Fourier mode ${\cal A}(X, \omega )$ for $\omega = \omega_0$
and $\omega = - \omega_0$ 
for $X >X_2$ and $X < X_1$ as described above.
Despite the scatter in the data it is clear that $\hat{D}(\hat{q})$ decreases with 
decreasing $\hat{q}$ and presumably goes to zero at a wavenumber for which the 
diffusion coefficient of the nonchaotic state is still negative as shown by the dashed line.
Thus the system is diffusively unstable over a range of wavenumbers, and initial conditions
with an average wavenumber in this range will evolve into domains of chaotic and nonchaotic
waves.
%The dashed line shows the presumed extension of the effective phase diffusion 
%coefficient necessary for our explanation of the observed localized chaos
%to be valid.

\begin{figure}[h]
\begin{picture}(420,180)(0,0)
\put(-50,-65) {\includegraphics{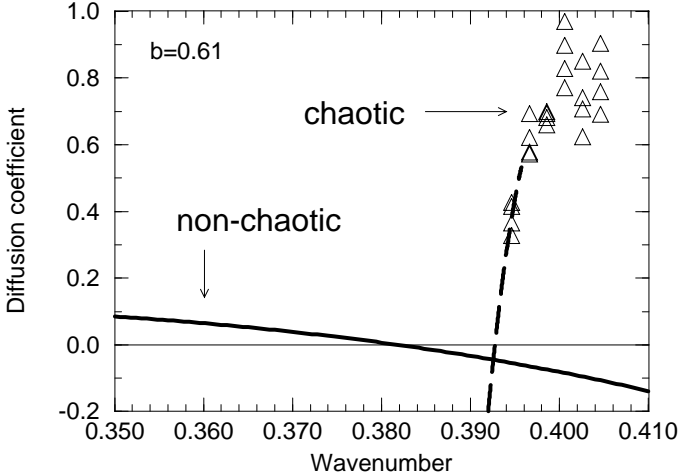}}
\end{picture}
\caption{Analytical phase diffusion coefficient for the stationary solution
(solid line) and effective diffusion coefficient for the chaotic solution
(triangles represent numerical results, the dashed line is the presumed
extension of these results).  ($b=0.61$, other parameters as in figure 1.)
\protect{\label{f:fig5}}
}
\end{figure}

Figure 5 suggests that the chaotic activity should not disappear homogeneously
if the average wavenumber is decreased towards the stable regime.
Instead, the homogeneously chaotic state should split up into chaotic and
stationary domains in which the local wavenumber is in the respective stable
regimes ($D>0$ and $\hat{D}>0$).
This is indeed found in our numerical simulations.

To summarize, we have presented numerical simulations of spatio-temporal chaos
in parametrically excited standing waves.
We have demonstrated that despite the fact that the chaos must be classified as
amplitude chaos its large-scale features can nevertheless be described by an 
effective phase equation.
The reason for this surprising result lies in the fact that the phase slips
always come in pairs such that the total phase is effectively conserved.
We have determined the effective diffusion coefficient.
This allowed us to identify a simple mechanism for the localization of the 
spatio-temporally chaotic state which is related to the occurence of domain
structures and zig-zags in steady patterns \cite{BrDe89,Ri90a,RaRi95a,BoKa90}.

So far we have not identified a particular physical system which exhibits the
behavior described here.
It is natural to suspect that the closing of the Eckhaus curve (cf. figure 1) is 
related to the occurence of double phase slips.
If that should be the case, parametric driving of waves which are Benjamin-Feir 
unstable in the absence of any forcing should be a good candidate for this 
behavior.
Within the framework of the coupled Ginzburg-Landau equations (1,2) 
these waves become unstable near the band 
center when forced sufficiently strongly \cite{Ri90}.
Depending on the sign of certain nonlinear coefficients this indicates 
instability at all wavenumbers and a closing of the Eckhaus curve.

We would like to thank J. Eggers for helpful discussions.
This work was supported by the United States Department of Energy through grant DE-FG02-92ER14303.

\bibliography{/home2/hermann/.index/journal}

\end{document}